# Are elite journals declining?

Vincent Larivière§, George A. Lozano†, and Yves Gingras*

§ *vincent.lariviere@umontreal.ca*
École de bibliothéconomie et des sciences de l'information, Université de Montréal, C.P. 6128, Succ. Centre-Ville, Montréal, QC. H3C 3J7 (Canada)
and
Observatoire des sciences et des technologies (OST), Centre interuniversitaire de recherche sur la science et la technologie (CIRST), Université du Québec à Montréal, CP 8888, Succ. Centre-Ville, Montréal, QC. H3C 3P8 (Canada)

† *dr.george.lozano@gmail.com*
Estonian Centre of Evolutionary Ecology, 15 Tähe Street, Tartu, Estonia, 50108

* *gingras.yves@uqam.ca*
Observatoire des sciences et des technologies (OST), Centre interuniversitaire de recherche sur la science et la technologie (CIRST), Université du Québec à Montréal, CP 8888, Succ. Centre-Ville, Montréal, QC. H3C 3P8 (Canada)

**Abstract**

Previous work indicates that over the past 20 years, the highest quality work have been published in an increasingly diverse and larger group of journals. In this paper we examine whether this diversification has also affected the handful of elite journals that are traditionally considered to be the best. We examine citation patterns over the past 40 years of 7 long-standing traditionally elite journals and 6 journals that have been increasing in importance over the past 20 years. To be among the top 5% or 1% cited papers, papers now need about twice as many citations as they did 40 years ago. Since the late 1980s and early 1990s elite journals have been publishing a decreasing proportion of these top cited papers. This also applies to the two journals that are typically considered as the top venues and often used as bibliometric indicators of "excellence", *Science* and *Nature*. On the other hand, several new and established journals are publishing an increasing proportion of most cited papers. These changes bring new challenges and opportunities for all parties. Journals can enact policies to increase or maintain their relative position in the journal hierarchy. Researchers now have the option to publish in more diverse venues knowing that their work can still reach the same audiences. Finally, evaluators and administrators need to know that although there will always be a certain prestige associated with publishing in "elite" journals, journal hierarchies are in constant flux so inclusion of journals into this group is not permanent.

**Introduction**

Since at least the middle of the 19th Century, scientific ideas and discoveries have been disseminated and discussed primarily via papers in journals (Harmon and Gross, 2007; Meadows, 1974). Until recently, these "papers" consisted literally, of paper, bound within issues and volumes, physically delivered on a regular basis to subscribing individuals and institutions. Since the 1990s, through the effects of internet, this age-old system has begun to be supplanted by the electronic dissemination of individual papers. In this new digital age, access to scientific papers has been radically transformed by web repositories of preprints (like ArXiv), institutional and personal repositories of accepted and published papers, electronic access of print journals, online-only journals and open access journals.

Russell and Rousseau (2002) noted that the Internet shifts the emphasis away from the journal and towards the individual article, and suggested that this trend could decrease the importance of the impact factor (IF). More specifically, given that in the digital age scientists have access to each paper individually without having to view the entire issue or volume of the journal, we predicted that the relationship between the IF and papers' citations would be weakening (Lozano, Larivière and Gingras, 2012). We tested that hypothesis using a dataset of over 29 million papers and 800 million citations, and we showed that from 1902 to 1990 the strength of the relationship between IFs and paper citations had increased, but as predicted, the variance of papers' citation rates around their respective journals' IF has been steadily increasing since 1991. Furthermore, until 1990, the proportion of top (i.e., most cited) papers published in the top (i.e., highest IF) journals had been increasing. However, since 1991 the pattern has reversed and we observe a decline in the proportion of top-cited papers in journals with highest IF. Hence, the most important literature is now published in increasingly diverse sources (Lozano, Larivière and Gingras, 2012).

However, the effect we documented at the macro level might not apply to the handful of "elite" journals that promote themselves as, and are perceived to be, the most important. Despite the changes in the publishing industry over the past two decades, this small group of elite journals could have kept publishing a stable or even increasingly larger proportion of the most cited papers. Here we compare citation rates in the 20 years before and after 1990, arbitrarily taken to be the time about which papers began to be disseminated digitally. Using the proportion of most cited papers by individual journals (Bornmann, Leydesdorff and Mutz, 2013; Waltman et al. 2012; Waltman and Schreiber, 2012), we examine the patterns found on a macro scale also occur in a selected number of elite journals, which are then compared with several journals that are becoming increasingly important.

**Methods**

We used Thomson Reuters' Web of Science (including all standard citation indexes, SCIE, SSCI and AHCI) from 1970 to 2010 (1970-2012 for citations received), with a total of more than 27.8 million papers and 784 million citations[1]. This database is based Thomson Reuters' source data transformed into a SQL relational database designed for bibliometric analyses. To have consistent measures over time, we used a fixed citation window of two-years following publication, to which we added citations received during publication year. We also compiled the data using a 5-year citation window following publication year, and the results were nearly identical. However, a 5-year citation window would eliminate the last 6 years of data, so only the 2-year citation data and analysis are presented. We used two citation thresholds: the top 5% most cited papers and the top 1% most cited. In both cases, citations received included self-citations. Given that the proportion of self-citations decreases with higher numbers of citations, self-citations are not a factor when one looks at the top cited papers (Glanzel et al., 2006). The papers included were articles, notes and reviews, and excluded editorials, as it is typically the case in bibliometric analyses (Moed, 2005). The citations they received could come from any type of document.

The calculation proceeded in 3 steps: 1) we calculated the number of citations each paper received; 2) we ranked the papers and selected the 1% and 5% top cited papers for each year; 3) we calculated the proportion of these papers published in the chosen set of journals. If that proportion declines for a given journal it means that its share of top-cited papers is decreasing, and vice-versa. It is important to distinguish this indicator, which tells us which journals obtain the larger part of top cited papers every year, from another indicator based on the proportion of citations that accrue over the years to papers published in those journals in a given year. Though the latter indicator has been used to measure a form of concentration of citation over time (Barabasi *et al.*, 2012), it is not a valid measure since it is affected by the half-life of citations received by papers. It is obvious that highly cited papers published in high impact journals will see their share of citations, within the specific journals in which they were published, accrue in the years following their publications. What we need to measure is whether or not the proportion of papers coming from elite journals and cited in the top 1% or 5% most cited papers is changing over the years.

We conducted this analysis for two groups of journals, which shall be referred as "elite" journals and "emerging" journals. "Elite" journals were chosen from the most prestigious journals, which also had the highest IFs in 2011 (amongst the top 1%). Additionally, these journals are not review journals, publish a high number of papers annually, and have been in existence for several decades. This group includes three general journals (*Nature*, *Science*, and *Proceedings of the National Academy of Science - PNAS*), and four medical/biomedical journals (*Cell*, *Lancet*, *New England Journal of Medicine - NEJM*, and the *Journal of the American Medical Association -* JAMA). *PNAS* has a slightly lower IF than the other journals, but was nonetheless included because of its reputation, and the fact that in the 1980s it published the highest proportion of highly cited papers. On the other hand, "emerging" journals were chosen from those that had the highest growth in their proportion of top cited papers over the past 40 years. This group included one general journal (*PLoS One*), three journals from material sciences/nanotechnology (*Nano Letters*, *Advanced Materials* and *Nature Materials*), one from medicine (*Journal of Clinical Oncology*) and one from chemistry (*Chemical Reviews*). Of these, *PLoS One*, *Nature Materials* and *Nano Letters* are relatively recent, created in this century, and the other three are older. *Chemical Reviews* was established in 1925, and the *Journal of Clinical Oncology* and *Advanced Materials* began publishing in the 1980s.

---

[1] Although references made by SSCI and AHCI journals are included, no journal from these two databases met the criteria to be considered as an elite journal.

Figure 1 presents the change of the percentage of all papers included in the top 1% and top 5% most cited (panel A) and the change of the citation threshold (panel B). The citation threshold for including papers in either the 1% or 5% category is an integer, but several papers can have this number of citations, so the percentage of papers included is always slightly above 1% or 5%. Citations are more spread out at the top tail of the citation distribution, so for the top 1% of papers, the number of papers included is actually closer to the actual 1% of papers. The number of citations needed to be included amongst the top cited papers has increased steadily (Fig. 1B). To be included among the top 5% or 1% most cited papers, in 2010 papers needed about twice as many citations as 30 or 40 years earlier.

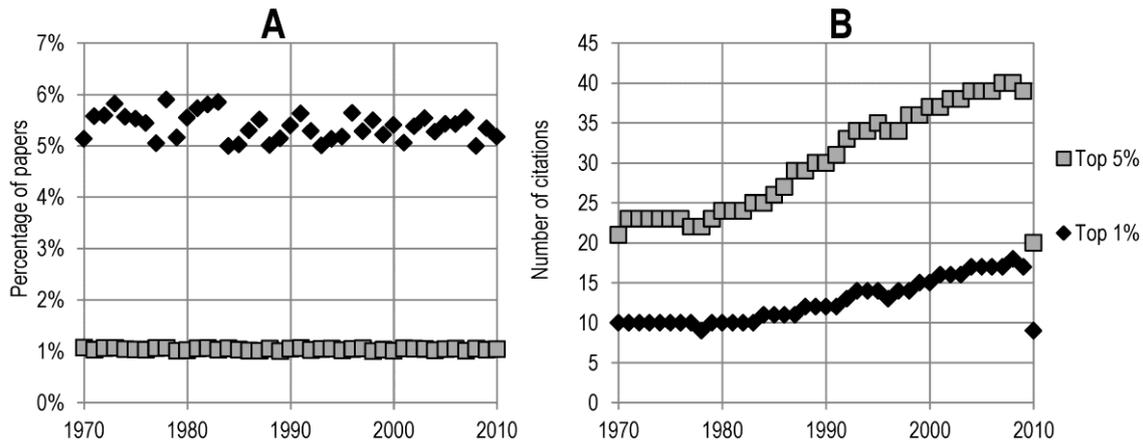

Figure 1. A) Proportion of papers in the top 5% and 1% most cited papers B) Citation threshold for inclusion in the top 5% and 1% most cited papers, 1970-2010.

**Results**

All seven elite journals are now publishing a smaller proportion of top cited papers than they did 20 to 25 years ago (Fig. 2). In all cases, these elite journals have been publishing a larger proportion of the top 1% most cited papers than of the top 5% most cited papers. The proportion of highly cited papers in the three general journals (*Science*, *Nature* and *PNAS*) and *Cell* had been increasing until around 1990 and decreased thereafter. In the mid-in1980s, *PNAS* was publishing almost 9% of the top1% most cited papers, and over 4% of the top 5% most cited papers. By 2010, these percentages have dropped considerably to about 2.7% and 2.2% respectively. Similarly, at their peak in the late 1980s and early 1990s *Nature* and *Science* published respectively about 7% and 6% of the top 1% most cited papers. These percentages are now about 4% and 3%, respectively. A similar trend is observed for the top 5% most cited papers, except that the increase in the first 20 years was evident only for *PNAS* and *Cell*. Among the biomedical journals, The *Lancet* and *NEJM*, follow a slightly different pattern when one considers the top 1% most cited papers. *NEJM*'s share has steadily decreased since the 1980s, and that of the *Lancet* has been relatively stable. However, both journals decreased their proportion of the top 5% most cited papers throughout the period, from about 1% in the 1970s to about 0.5% in 2010. Finally, *JAMA* increased its share of the top 5% most cited until the second half of the 1990s and of the top 1% cited papers until the early 2000s, and slowly lost ground afterwards.

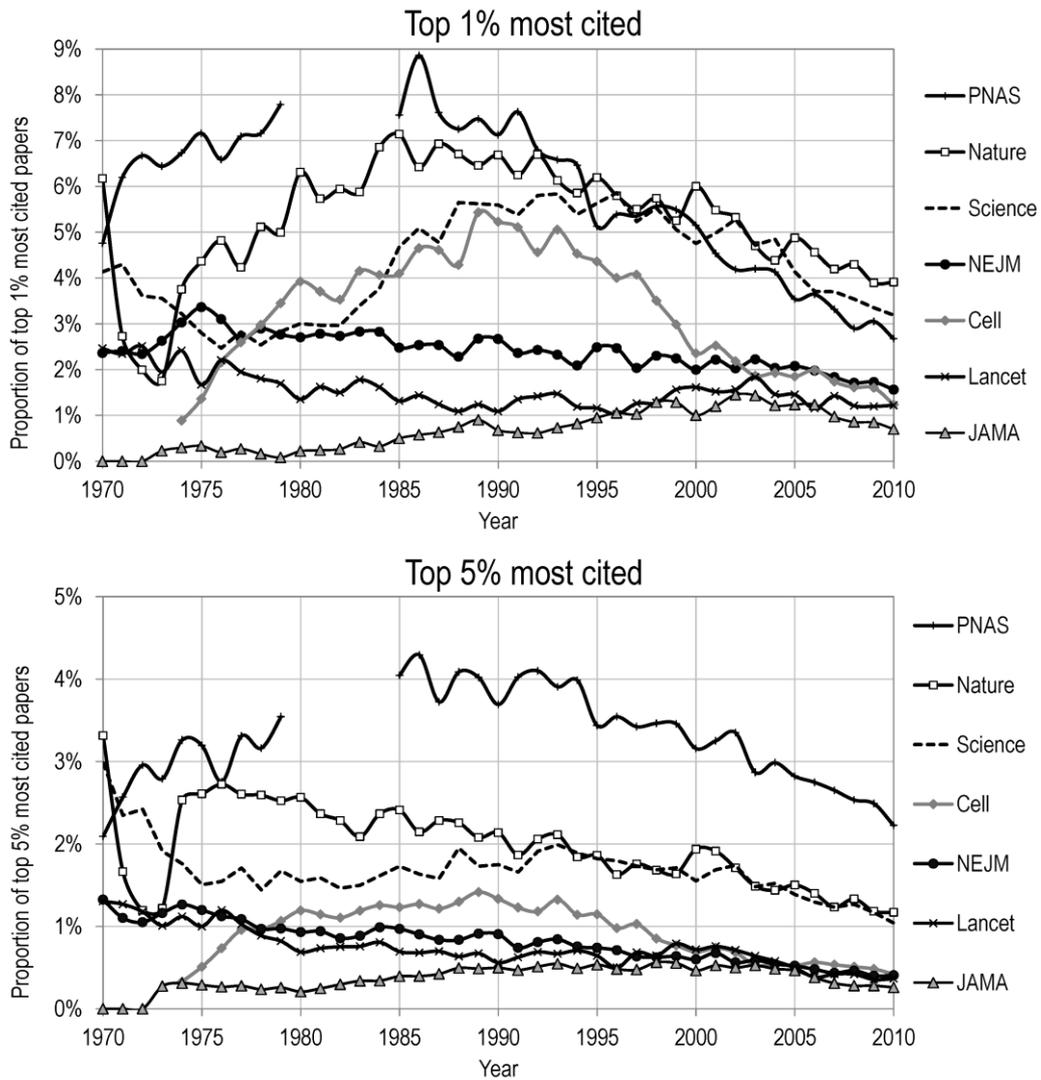

Figure 2. Proportion of top 1% and 5% most cited papers in elite journals, 1970-2010. *PNAS* data were not available from 1980 to 1984.

Emerging journals were selected because they have been increasing their share of highly cited papers (Fig. 3). Of the 3 newer journals, *PLoS One*, an open access interdisciplinary journal founded in 2006, currently accounts for 0.6% of the top 1% most cited papers and 0.8% of the top 5% most cited papers. Similarly, *Nature Materials* and *Nano Letters* currently each have over 0.5 % and 1% of the top 1% most cited papers, respectively. The three older journals, *Chemical Reviews*, *Advanced Materials* and the *Journal of Clinical Oncology* have managed to increase their proportion of top cited papers since the early 1980s.

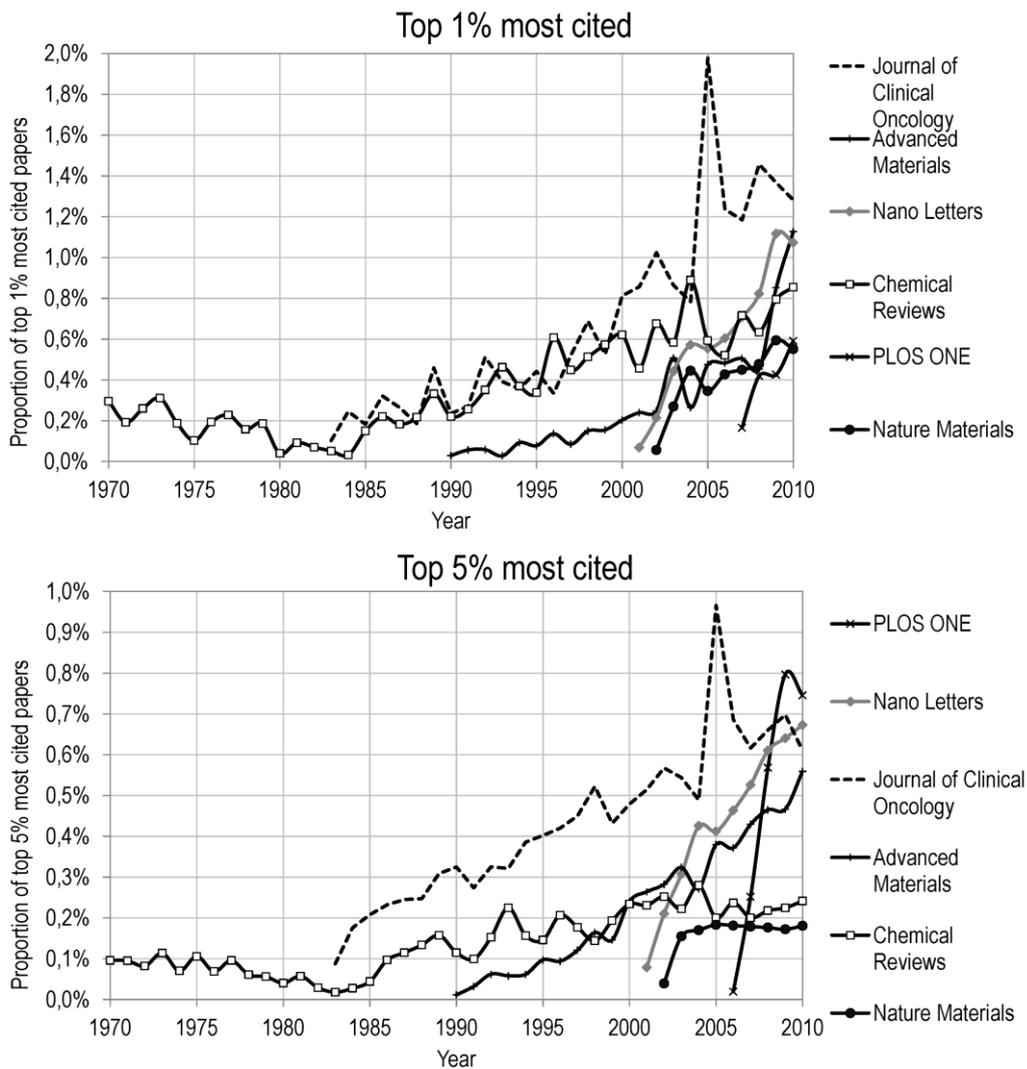

Figure 3. Proportion of top 1% and 5% most cited papers for emerging journals, 1970-2010.

The total number of papers published has been increasing during the past several decades, and in most cases, the number of papers published yearly by individual journals has also been increasing. To assess whether the patterns documented above were simply due to increases in the number of papers published by each journal relative to the total number of papers published, we computed a yearly normalized top 1% index for each journal. The normalized top index is a measure of the number of top papers (1%, 5%, etc., here we use the 1% threshold) published in a journal in a given year relative to what would be expected if the top papers were randomly distributed throughout all journals. More specifically, for a given year, it is the relative number of top papers in a given journal (top papers / total papers) divided by the proportion of top papers published in all journals. A coefficient of 1 would indicate that the number of top papers published by a journal is what would be expected by chance. A coefficient of 10 indicates that a journal published 10 times as many top papers as would be expected by mere chance. For example, if one million papers are published in a given year, 10,000 would be in the top 1%, and if a given journal publishes 500 papers that year, and 25 are among these 10,000, then its normalized top 1% index would be 5."

Elite journals publish far more top papers than would be expected by chance, but different patterns occur (Fig. 4A). In the first 20 years *Cell* went from about 30 to 70, and today it is back down to about 50. From 1970 to 1990, *JAMA* and *Lancet* were in the single digits and low teens, respectively, and then increased sharply to the 40 to 50 range. *PNAS* held steady at the low 20s until about 1990 and since then has been slowly decreasing to about 10. Finally, in the past 20 years *Nature*, *Science* and *NEJM* have been slowly increasing, but still more or less within the same range, from about 40 to about 50. Interestingly, *Nature* and *Science* had their sharpest increases in the 1970s and 1980s from 10 to about 40. These patterns do not necessarily differ along subject lines.

Emerging journals are also publishing a higher proportion of top 1% most cited papers than expected by chance given the number of papers they publish (Fig. 4B). *Chemical Reviews* and *Nature Materials* are now in the 40 to 50 range. Oddly, from 1970 to 1990 *Chemical Reviews* was highly variable. *Journal of Clinical Oncology*, *Advanced Materials* and *Nano Letters* are now in the high teens. For *PLoS One*, this ratio has remained in the low single digits, and is only at 1.09 in 2010, which shows that its increase in the proportion of top cited papers (Fig. 3B) is mostly due to the very high number of papers it publishes.

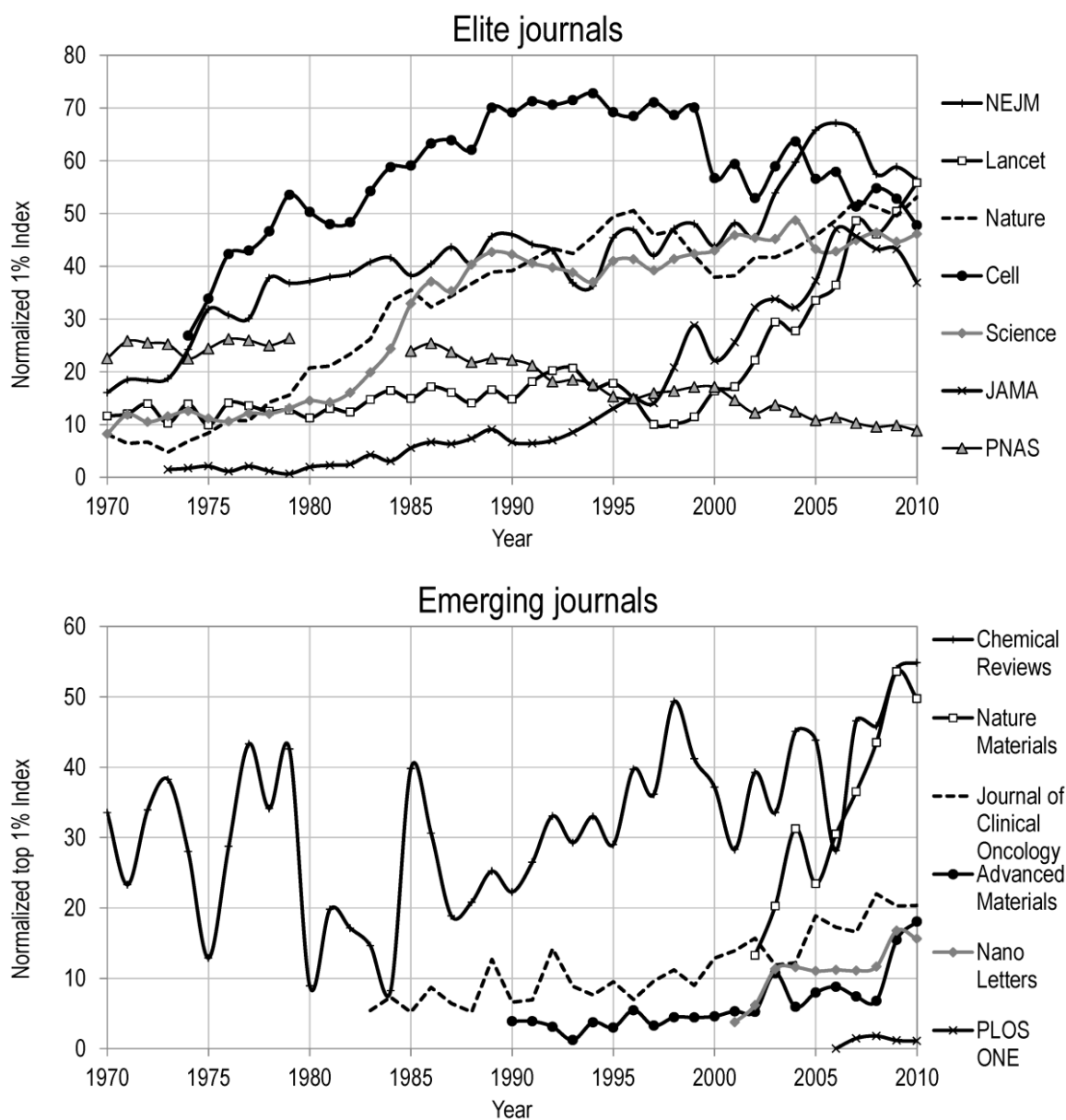

Figure 4. Normalized 1% Index for elite and emerging journals, 1970-2010.

**Discussion and Conclusion**

Previously, in a large scale analysis, we documented that the proportion of the most cited papers published in the highest IF journals had been steadily decreasing since the advent of the digital age (Lozano, Larivière & Gingras, 2012). Here, we tested whether this pattern was also true for the handful of elite journals that are generally considered the most important. Our preliminary analysis indicated that the definition of "highly cited" has changed. More papers are published now and these papers have longer references lists, so papers receive more citations and fewer papers go completely uncited (Wallace, Larivière & Gingras, 2009). Consequently, to be among the top nth percentile, papers now need about twice as many citations as they did 30 to 40 years ago. The effect of this inflation is such that citations have devalued by about half in a time span that is within the range of a scientific career, which might affect the interpretation of career-long citation-based impact measures for individual researchers.

Since the late 1980's and early 1990s, several new and some long-established journals are becoming more important, while traditional elite journals, including *Science* and *Nature,* are publishing a decreasing proportion of the top cited papers. However, even though their share of the most cited articles is declining, elite journals still "punch above their weight" and publish a larger proportion of top cited papers than we might expect from their total number of papers. This relative decline of elite journals amongst all journals is in fact consistent with the general trend towards the deconcentration of research activities. At the level of countries as well as within countries, the production of papers is more and more diversified geographically and thus less concentrated in a few top countries or cities (Grossetti *et al.*, 2012). In terms of citations received, the distribution is less concentrated now than in the past, as more papers get their share of citations (Larivière, Gingras and Archambault (2009).

The decline of elite journals was caused by the internet via several inter-related effects. First, the digital age made papers more independent of their respective journals. The digital age has transformed the manner in which scientists find papers. We used to search for papers on the bookshelves of the library but now we do so via the internet, often at our library's website. Furthermore, papers are now directly accessible independently and one does not have to even look at the corresponding issue or volume of the journal. Hence, whether papers get cited or ignored is increasingly independent of the journal in which they appear.

Second, the digital age also facilitated the creation of new journals. Several high quality journals were triggered by the creation of new fields (e.g., *Nano Letters*, *Advanced Materials*) but often it was simply because digital journals are easier and cheaper to produce and distribute than print-based journals. As the journals were created, journals that were already in the system were likely to lose market share, and the ones more likely to be negatively affected were those having the highest proportion of top papers. A similar phenomenon occurred for the share of papers at the level of countries. When China increased its share of the world's research papers during the 1990s and 2000s, the share of the most important countries declined, as was the case for United States and Japan (Leydesdorff and Wagner, 2009; Zhou & Leydesdorff. 2006). Dozens of new publishers and hundreds of new journals have appeared in the past 20 years. Inevitably, elite journals had to lose some of the top papers to these new competitors.

Third, the creation of some of these new journals was motivated by ideals of "free access to all" that took advantage of the ease with which information can be disseminated in the digital age (e.g., *PLoS One*). Papers that are freely accessible have a greater probability of being cited (Gagouri *et al.*, 2010). Researchers now access papers from a greater variety of journals, not just the so-called premier journals in a given field, or elite journals in general. Technically, a paper could now be in any journal, and it would still be found by internet searches, downloaded if available, and cited if deemed relevant.

Although elite journals were generally declining, the different patterns we documented do not fall neatly along subject lines, general vs. biomedical. Similarly, the patterns observed in emerging journals do not depend on the field of study, or the age or novelty of the journal. Among emerging journals, *PLoS One*, the only one examined here that is completely "open", had a high proportion of highly cited papers, but only because it publishes a high number of papers. Hence, on average, its papers are not necessarily better but at least they can be found and read by anyone, free of charge. This complexity over the past 20 years indicates that the digital age affected each journal differently, likely depending on the effectiveness of their editorial, advertising and marketing policies.

Elite journals have been aware of the threat of new journals and have tried to counter their effect. For example, Nature Publishing Group, a private company, instead of just expanding the flagship journal, started to create specialized journals in the 1990s (*Nature Genetics, Nature NeuroScience, Nature Medicine*, etc.) and review journals in the 2000s (*Nature Reviews Cancer, Nature Reviews Immunology*, etc.). They now publish 37 other journals with the name "Nature". These new journals capitalized on the symbolic capital of the original journal, *Nature*, and they rapidly attracted high impact papers. One potential explanation for choosing this route, a phenomenon that economists call *versioning*, is that the publisher can charge new subscription and advertising fees for the new journals, whereas it might be difficult to make consumers and advertisers accept large price increases for the flagship journal even when accompanied by proportional increases in the size of the journal. These new journals spread and diluted the impact of the flagship journal across all of the publisher's journals. Combining all of Nature Publishing Group journals, the proportion of the top 5% papers increased from 3% in 1970 to 5% in 2000, and has remained stable since then. On the other hand, the journal *Science* is published by the American Association for the Advancement of Science (AAAS), a non-profit organization that publishes only 3 peer-reviewed journals.

With all these new journals, elite and otherwise, researchers now have increasingly more venues where they can submit their papers, and benefit from a visibility and availability that formerly was possible only for the most widely distributed journals, the "elite" journals. Given the high rejection rates of elite journals, around 93% for *Science*, for example, researchers might prefer to save time and submit their papers to other journals that ultimately will reach the same audience faster and potentially obtain as many citations. In the digital age it is relatively easy to determine the actual citation rate of individual papers or authors, so the value of a journal's reputation is now less important. Nevertheless, researchers might still prefer to publish in elite journals. Whether justifiable or not, journal reputation still has some value to the papers therein, through a Mathew effect (Larivière and Gingras 2010), particularly when the research is viewed and evaluated by non-experts.

Finally, advancement, recruitment and grant-evaluation committees, administrators and other evaluators should heed these results. The quality of papers and competence of researchers should be evaluated independently of the journal in which the work appeared for two reasons. First, traditional "elite" journals still have the highest citation impact, but other journals are also publishing an increasingly higher proportion of top cited papers. Second, even if this journal hierarchy were valid, the hierarchy is not fixed. In fact, even if journal reputations are loosely based on a journal's tendency to publish high quality work, reputations have an intrinsic inertia and there is always a time lag between a journal's actual value and its reputation. Although there shall always be a hierarchy of journal prestige, the hierarchy is dynamic. Many other, more suitable criteria can be used to assess researchers, but if journal quality must be used, equity demands that evaluators become attentive to changes in the scientific publishing landscape.

**Acknowledgements**

We thank Stefanie Haustein and Jean-Pierre Robitaille for stimulating discussions on the topic. G. Lozano thanks the University of Tartu for unhindered and free access to their online collections.